## Murray Moinester

School of Physics and Astronomy, Tel Aviv University, 69978 Tel Aviv, Israel
email: murray.moinester@gmail.com

# Tests of Three-Flavor Chiral Perturbation Theory

## Abstract

We review experimental tests of three-flavor (*u,d,s*) chiral perturbation theory (ChPT). These include measurements of pion and kaon polarizabilities, the chiral anomaly amplitudes for processes such as $\gamma\rightarrow\pi\pi\pi$, $\gamma\rightarrow\pi\pi\eta$ and $\gamma\rightarrow KK\pi$, as well as the lifetimes of the neutral pion and eta. These observables are extracted primarily through ultra-peripheral Primakoff scattering of high-energy particles from virtual photons in the Coulomb field of nuclei. Comparing data to two-flavor and three-flavor predictions allows us to evaluate how well ChPT describes light meson dynamics and the role of the strange quark in spontaneous chiral symmetry breaking.

## 1. Introduction:

Consider the amplitude for the forward high-energy scattering of a beam particle off a nuclear target with atomic number Z, mediated by one-photon exchange. The photon exchange is proportional to the fine-structure constant $\alpha$ and inversely proportional to the squared four-momentum transfer t. The value of t is itself inversely proportional to the squared four-momentum center-of-mass energy s, and decreases rapidly as s increases. Despite the weak electromagnetic interaction, the interaction amplitude can therefore still be significant.[1] Moreover, the very small value of t allows experiments to distinguish between ultra-peripheral Coulomb scattering and nuclear background interactions involving larger momentum transfers. In addition, the nuclear form factor at very low t is at its maximum, so it does not suppress the scattering amplitude. Using the Weizsäcker-Williams equivalent photon approximation[2,3,4] to describe the quasireal photons of the Coulomb field of the target enables the extraction of low-energy quantities that can be compared with predictions from chiral perturbation theory (ChPT). Fermi first suggested this approximation, noting that the electromagnetic fields of a rapidly moving charged particle resemble those of a photon pulse.[5] Here, the spherically symmetric Coulomb electromagnetic field of the target nucleus looks like a transverse photon pulse when viewed from the rest frame of the incident beam particle.

Pion polarizability[1,6,7] and chiral anomaly $\gamma\rightarrow\pi\pi\pi$[8,9] measurements were performed at CERN COMPASS[10,11] using high energy pion scattering from high-Z nuclei. We discuss the good agreement of these studies with 2-flavor ChPT predictions. CERN AMBER[12] plans to study similar kaon-induced reactions, following upgrading the COMPASS beamline[13,14] by setting up a high-energy and high-intensity kaon beam. These reactions include the kaon polarizability, $\gamma\rightarrow\pi\pi\eta$, and $\gamma\rightarrow KK\pi$ chiral anomalies.[15,16,17] At Jefferson Laboratory (JLab), $\pi^0$ lifetime measurements were performed by the PrimEx collaboration using γ Primakoff scattering from high-Z nuclei.[18] PrimEx plans similar measurements of the η lifetime[19,20] and pion polarizability.[21] All these studies are important for checking the theoretical framework of 3-flavor ChPT.

## 2. Primakoff Effect and Scattering

The Primakoff effect, $\gamma\gamma^* \rightarrow$ meson, describes the resonant production of neutral pseudoscalar mesons ($\pi^0$ or η) through the interaction of high-energy photons (γ) with quasi-real photons ($\gamma*$) of the Coulomb field of a nucleus.[22,23,24] It is equivalent to the inverse kinematics process where a $\pi^0$ or η decays into two photons. It has therefore been utilized to measure the decay width (lifetime) of neutral mesons. Relatedly, Primakoff scattering[25] involves high-energy beam particles (charged or photons) scattering off quasireal photons ($\gamma*$) within the Coulomb field of a nucleus. This process is referred to as Bremstrahlung when high-energy electrons are the beam particles, with the scattered electrons and real photons projected into forward angles in the laboratory. In Primakoff scattering, quasireal photons ($\gamma*$) act as a target having effective thickness proportional to $Z^2$. The present review focuses on Primakoff effect and scattering data.

## 3. The Chiral Perturbation Theory (ChPT) Lagrangian

Quantum Chromodynamics (QCD) is a quantum field theory describing the strong interactions of quarks via the exchange of massless spin-1 gauge bosons, called gluons, which carry color charge. Perturbative QCD predictions of hard phenomena have been experimentally confirmed.[26,27] However, for low-energy interactions, non-perturbative effects become significant. In QCD, in the limit of vanishing quark masses ($m_q \rightarrow 0$), the left- and right-handed quarks, denoted as $q_L$ and $q_R$, decouple. The effective QCD Lagrangian exhibits an $SU(3)_L \times SU(3)_R$ chiral symmetry. Chirality (handedness) is based on the relative orientations of spin and momentum. The index q denotes the quark flavors *u, d,* and *s*. Chiral symmetry refers to the invariance of QCD under transformations that act separately on left-handed and right-handed components of the quark fields. While the Lagrangian for $m_q \rightarrow 0$ possesses chiral symmetry, the vacuum state, the lowest-energy state, does not. Chiral symmetry breaks spontaneously in the vacuum state. In addition, chiral symmetry is explicitly broken by the small masses of bare quarks.

For hard processes, at separations less than approximately 0.2 femtometers, quarks and gluons behave as nearly free particles due to a phenomenon known as asymptotic freedom. The intermediate distance scale, roughly 0.2 to 0.5 fm, marks the region where quarks and gluons are confined within hadrons and where chiral symmetry becomes spontaneously broken. Both phenomena are fundamental to the transition from the perturbative, asymptotically free regime at short distances to the non-perturbative regime that governs hadron structure. At distances greater than about 0.6 fm, non-perturbative effects, such as quark condensates and hadronization, dominate the dynamics.

QCD calculations at low momentum transfers may be performed using the effective field theory formulated by Weinberg[28] and systematically developed by Gasser and Leutwyler.[29] It incorporates a systematic expansion at low momenta, known as Chiral Perturbation Theory (ChPT), applicable in the non-perturbative low-energy regime of strong interactions.[30,31] ChPT employs an effective Lagrangian $L_{eff}$ which is expressed as a power series in pion momenta and mass terms, capturing the low-energy dynamics. The ChPT Lagrangian encodes how the pions interact. The derivatives in the Lagrangian represent how the fields change in space and time—essentially, their momenta.

The Lagrangian $L_{eff}$ is derived by considering the effects of spontaneous symmetry breaking and integrating out the high-energy dynamics of the quarks and gluons. $L_{eff}$ embodies Goldstone's theorem[32,33], which states that for every spontaneously broken continuous symmetry, there exists a corresponding massless pseudoscalar particle known as a Goldstone boson.[26] The three pions ($\pi^+$, $\pi^-$, $\pi^0$) play central roles as (approximate) Goldstone bosons when ChPT is restricted to the *u* and *d* flavors. The eight lowest mass $J^P = 0^-$ mesons ($\pi$o, $\pi^{\pm}$, $K^{\pm}$, $K^0$, $\overline{K}^0$, η) emerge as the corresponding Goldstone bosons for 3-flavor (u,d,s) ChPT. Leff successfully describes the interactions of light mesons at low energies, such as the $\pi\pi$ scattering length.[28] Summarizing, ChPT describes the interactions of pions, kaons, and etas at energies below 1 GeV, where quark and gluon degrees of freedom are confined, and the relevant degrees of freedom are the Goldstone bosons arising from spontaneous chiral symmetry breaking.

The QCD vacuum is not empty; rather, it possesses a highly non-trivial structure characterized by a non-zero quark condensate, which acts as an order parameter for spontaneous chiral symmetry breaking.[26,27] The "chiral condensate" quantifies the vacuum expectation value of quark-antiquark pairs, $<0|q\,\bar{q}|0>$, reflecting the degree of vacuum polarization. Gell-Mann, Oakes and Renner[34] demonstrated explicitly that the squared pion mass is proportional to the average light quark (*u, d*) mass times the magnitude of the chiral condensate. The agreement between the observed pion mass and this relation supports the view that the vacuum condensate's magnitude fundamentally governs the pattern of chiral symmetry breaking. This connection between the vacuum structure and low-energy observable properties provides a strong theoretical foundation and justifies the use of ChPT, which exploits the spontaneously broken symmetry and effective Lagrangian framework to describe mesonic phenomena at low energies.

Here, we focus on processes involving photons and mesons. Charge gauge invariance enables incorporating photons into $L_{eff}$ by introducing electromagnetic coupling terms that are consistent with chiral symmetry. ChPT provides rigorous predictions of $\gamma\pi$ interactions at low energies, stemming directly from QCD and relying on spontaneously broken chiral symmetry, Lorentz invariance, and low momentum transfer. Unitarity is achieved by incorporating pion-loop corrections, the resulting infinite divergences of which are absorbed into the physical (renormalized) coupling constants $L_r$. By restricting the perturbative expansion of $L_{eff}$ to quartic order in momenta and masses, $O(p^4)$, this method establishes connections between various processes through a common set of 12 $L_r$ constants.[29,30,31]

The pion polarizabilities discussed below are related by two of the $L_r$ constants to the ratio $h_A/h_V$ measured by radiative pion beta decay, $\pi^+ \rightarrow e^+ \nu_e \gamma$, where $h_A$ and $h_V$ are the axial vector and vector coupling constants that characterize the decay, respectively.[1,35] ChPT is anticipated to account for the dominant portion of the cross-sections in low-momentum transfer reactions involving pions, kaons and photons, particularly in the Primakoff reactions related to polarizability, chiral anomaly, and lifetime experiments that are studied at CERN COMPASS, CERN AMBER and JLab.

## 4. Gamma-Pion Compton Scattering and Pion Polarizabilities

The $\gamma\pi \to \gamma\pi$ Compton scattering cross section depends predominantly on the pion charge. In the pion rest frame, for $\gamma$ photon energies in the range of $\hbar\omega$ = 60−780 MeV and scattering angles greater than 80°, approximately 5% of the differential cross-section depends on the electric $\alpha_\pi$ and magnetic $\beta_\pi$ charged pion polarizabilities.[6] These moments are induced via the interaction of the photon's electromagnetic field with the quark substructure of the pion.[1,35] In particular, $\alpha_\pi$ is the proportionality constant between the electric field of the photon and the induced electric dipole moment, whereas $\beta_\pi$ is the proportionality constant between the magnetic field of the photon and the induced magnetic dipole moment. Polarizabilities are fundamental pion characteristics.

The polarizabilities may be extracted from the shape of the $\gamma\pi$ Compton scattering differential cross-sections in the pion rest frame.[1,6] The influence of the polarizabilities increases with increasing photon energy and photon scattering angle. COMPASS measured $\gamma\pi$ scattering via Primakoff scattering with 190 GeV negative pions, $\pi Z \to \pi Z \gamma$, where Z is the nuclear charge, and Ni was the target. In the one-photon exchange domain, this is equivalent to $\gamma\pi \to \gamma\pi$ scattering. Assuming $\alpha_\pi+\beta_\pi=0$, the dependence of the laboratory differential cross-section on $x_\gamma=E_\gamma/E_\pi$ was utilized to determine $\alpha_\pi$, where $x_\gamma$ is the fraction of the beam energy carried by the final state $\gamma$.[1,6] The variable $x_\gamma$ is related to the photon scattering angle for $\gamma\pi \to \gamma\pi$.

## 5. Future Polarizability Studies

**JLab:** Experiment E12-13-008[21] at Jefferson Laboratory plans to measure $\gamma\gamma \to \pi^+\pi^-$ cross-sections and asymmetries via Primakoff scattering using a 6 GeV linearly polarized photon beam, a Sn "photon target" and the JLab GlueX detector. The expected azimuthal angle asymmetry dependence will be used to reduce the background. They aim to achieve an approximate 10% uncertainty for $\alpha_\pi - \beta_\pi$.
**COMPASS:** The COMPASS experiment at CERN has already obtained approximately 5 times higher statistics data for the pion polarizability compared to the published result.[1,7] This is expected to improve the determination of $\alpha_\pi-\beta_\pi$, and may possibly even allow a measurement of $\alpha_\pi+\beta_\pi$.
**AMBER**: The AMBER experiment at CERN aims to measure kaon polarizabilities via kaon Primakoff scattering.[14,17]

## 6. Polarizability Comparison between Experiment and Theory

Testing ChPT is possible by comparing the experimental polarizabilities with the ChPT low order (LO) and high order (HO) predictions of $\alpha_\pi - \beta_\pi = (5.4\pm1.0)\times10^{-4}$ fm$^3$ and $\alpha_\pi - \beta_\pi = (5.7\pm1.0)\times10^{-4}$ fm$^3$ , respectively.[1,35,36] Considering uncertainties, the COMPASS measurement[7], $\alpha_\pi-\beta_\pi = (4.0\pm1.2_{stat}\pm1.4_{syst})\times10^{-4}$ fm$^3$, agrees with both LO and HO predictions. Future high-precision Primakoff experiments are planned at CERN and Jefferson Lab. At CERN, these include the AMBER projects for kaon polarizabilities[14,17] alongside continued analysis of existing CERN COMPASS pion polarizability data. At JLab, a pion polarizability measurement is planned.[21] Combined with further theoretical developments, these data should provide important input for comprehensive tests of three-flavor ChPT.

## 7. Chiral Axial Anomaly

In the $\gamma\pi$ interaction at order $p^4$, the effective Lagrangian $L_{eff}$ must include Wess-Zumino-Witten (WZW) terms.[36,37,38,39] These are essential because quantum anomalies— specifically, the non-zero divergence of the axial current—arise from triangle diagrams involving quark loops in QCD. The WZW terms encode this structure within the effective mesonic framework, ensuring that the required anomalous Ward identities[40] are satisfied. These guarantee that the classical conservation law, the conservation of the axial current, is broken at the quantum level by the anomaly.

By including the WZW action, $L_{eff}$ accurately reproduces the effects of the anomaly, which manifest in processes such as $\pi^0 \to \gamma\gamma$, $\gamma\pi \to \pi\pi^0$, $\gamma\pi \to \pi\eta$[15] and $\gamma K \to \pi^0 K$.[16] In particular, the anomalous processes $\pi^0 \to \gamma\gamma$ and $\gamma \to \pi\pi\pi$ are characterized by amplitudes $A_\pi$ and $A_{3\pi}$ that measure their strength. Precise experimental determinations of these amplitudes are crucial for testing the chiral anomaly predictions. Theoretical approaches based on lattice QCD and dispersion analyses provide complementary descriptions of the chiral anomaly.[41]

## 8. The $\pi^0 \to \gamma\gamma$ $A_\pi$ Amplitude

The chiral anomaly amplitude $A_\pi$ is a crucial indicator of the interaction strength of pions, directly influencing the decay width Γ and mean lifetime τ of $\pi^0$. These quantities are related through the Heisenberg Uncertainty Principle by $2(\Delta E)(\Delta t)=\hbar$, or $\Gamma\tau=\hbar$; where $\Gamma=2\Delta E$ is the full width at half maximum of the resonance peak in the experimentally measured $\pi^0$ mass distribution, and Δt is the total time (lifetime τ) available for the mass measurement. This yields $\Gamma\tau=65.82\text{x}10^{-17}$, with Γ in eV and τ in seconds. Since $\Gamma(\pi^0\to\gamma\gamma) = BR(\pi^0\to\gamma\gamma)\,\Gamma(\pi^0)$, where the Branching Ratio is BR=0.988, the $\pi^0$ lifetime is calculated using $\Gamma(\pi^0)$ rather than $\Gamma(\pi^0\to\gamma\gamma)$.

The ChPT prediction at $O(p^4)$ is given as $A_\pi = \alpha/\pi f = 0.0252$ GeV$^{-1}$, where $\alpha$ is the fine structure constant and f=92.4 MeV is the pion decay constant[42], slightly different from the PDG value f=92.3 MeV.[43] The decay width is given[44] via ChPT as $\Gamma(\pi^0) = \alpha^2 m^3/64\pi^3 f^2$. Reviews of $\pi^0$ lifetime calculations and experiments include those by Bernstein and Holstein[44] and Miskimen.[45]

Table 1 summarizes the theoretical and experimental results for the $\pi^0 \to 2\gamma$ decay, combining statistical and systematic errors in quadrature when the two are given. The most recent and precise measurement of the $\pi^0$ lifetime by the Primakoff effect was conducted at JLab by the PrimEx collaboration, with the result[18] $\tau(\pi^0) = (8.34 \pm 0.06 \text{ stat.} \pm 0.11 \text{ syst.})\times 10^{-17}$s. This disagrees with a direct lifetime measurement[46] based on the $\pi^0 \to 2\gamma$ mean decay distance, $\tau(\pi^0) = (8.97 \pm 0.25) \times 10^{-17}$s. An estimated average $\pi^0$ momentum was used in this direct lifetime evaluation. A possible reason for the disagreement is that the error in $\tau(\pi^0)$ is larger than the 3% estimated error.[44,47] In the case of the $e^+e^-\to e^+e^-\gamma\gamma$ lifetime result[48], $\tau(\pi^0) = (8.4 \pm 0.8) \times 10^{-17}$s, the 9% uncertainty is too large to validate theoretical calculations. Thus, we focus on the 1.5% PrimEx experimental result.

Three high-order (HO) 3-flavor ChPT lifetime calculations were carried out.[47,49,50] Since they are approximately equal, we show only the two-loop HO calculation[50] alongside the 2-flavor LO theory results.[44] The HO result for $\tau(\pi^0)$ is 4.0% lower than the LO value, which was attributed to HO corrections involving isospin breaking and mixing small η and η′ components into the $\pi^0$ wave function.[47] The PrimEx result agrees well with the lowest order (LO) two-flavor (*c, d*) chiral anomaly prediction, confirming the applicability of LO 2-flavor ChPT in the chiral anomaly sector. Although the HO result is 2.4% lower than the PrimEx value, they are consistent considering uncertainties. Such a difference with smaller uncertainties would suggest limitations in 3-flavor ChPT.

Table 1. Theoretical and experimental results for the $\pi^0 \to 2\gamma$ decay

| Theory(T), Exp(E) | Reference | $\Gamma(\pi^0\to\gamma\gamma)$ eV | $\Gamma(\pi^0)$ eV | $\tau(\pi^0)$ ($10^{-17}$ s) |
|---|---|---|---|---|
| T LO | 44 | | 7.76±0.02 | 8.48±0.02 |
| T HO | 50 | | 8.09±0.11 | 8.14±0.11 |
| E, PrimEx | 18 | 7.80±0.12 | 7.90±0.12 | 8.34±0.13 |
| E, Direct | 46 | | 7.34±0.21 | 8.97±0.25 |
| E, $e^+e^-$ | 48 | 7.7±0.7 | 7.8±0.7 | 8.4±0.8 |

## 9. The $\gamma\to\pi\pi\pi$ $A_{3\pi}$ Amplitude

Within leading-order (LO) ChPT, the $A_{3\pi}$ chiral anomaly amplitude is given by $A_{3\pi} = e/(4\pi^2\mathbf{f}^3) = 9.72$ GeV$^{-3}$, where $e = \sqrt{4\pi\alpha} = 0.3028$ and **f** is the pion decay constant.[8,42] Bijnens *et al.*[51] examined higher-order ChPT corrections in this abnormal intrinsic parity sector. They included one-loop diagrams involving one vertex from the WZW term, and tree diagrams from the $O(p^6)$ Lagrangian. The parameters of the Lagrangian were estimated using vector meson dominance calculations. For $A_{3\pi}$, these HO corrections increase the LO value by approximately 10% to a value estimated here as $A_{3\pi} = 10.7 \pm 1.0$ GeV$^{-3}$.

The $A_{3\pi}$ amplitude was first determined by Antipov *et al.* at Serpukhov via the Primakoff reaction, $\pi^- Z \to \pi^-\pi^0 Z'$, using 40 GeV pions.[52] In the one-photon exchange domain, this corresponds to $\gamma\,\pi^- \to \pi^0\,\pi^-$, whose cross section is proportional to $(A_{3\pi})^2$. They found $A_{3\pi} = 12.9 \pm 1.0$ GeV$^{-3}$, with statistical and systematic errors added here in quadrature. Ametller *et al.*[53], taking electromagnetic contributions into account, reduced this value to $A_{3\pi} = 10.7 \pm 1.2$ GeV$^{-3}$. These data were also independently reanalyzed by Holstein[54], Hannah[55] and Truong.[56] They found that taking momentum dependence into account reduces the experimental $A_{3\pi}$ value by ~10%. Therefore, for the purpose of this review, the Serpukhov "value" is taken to be[52-56] $A_{3\pi} = 9.6 \pm 1.2$ GeV$^{-3}$.

$A_{3\pi}$ was also determined from CERN data of high-energy pion scattering from electrons in $H_2$ atomic orbits, $\pi^- e \to \pi^- e\, \pi^0$. Amendolia *et al.*[57] reported 36 events for this reaction, corresponding to a cross-section of 2.11 ± 0.47 nb. Giller *et al.*[58] carried out Monte Carlo integrations of the cross-section within the kinematic range of the experiment using different theoretical expressions for $A_{3\pi}$. By using an $O(p^6)$ ChPT $A_{3\pi}$ amplitude expressio, including electromagnetic corrections, taking into account momentum transfer dependence, and comparing the integrated cross-section to the data, they obtained $A_{3\pi} = 9.6 \pm 1.1$ $GeV^{-3}$. However, the uncertainties in this value are too large for ChPT testing.

Consequently, the discussion here focuses on the $A_{3\pi}$ amplitude, most recently measured by the COMPASS collaboration through Primakoff scattering.[9] This process, involving the chiral anomaly $\pi\gamma \to \pi\pi$, as well as the radiative ρ production reaction $\pi\gamma \to \rho \to \pi\pi$, both lead to $\pi^-\pi^0$ states. COMPASS measured the $\pi^-\pi^0$ Primakoff production cross-section starting from the kinematic threshold, where the chiral anomaly dominates, up through the ρ(770) resonance region. Data analysis allowed for the extraction of both the chiral anomaly amplitude $A_{3\pi}$, and the radiative width of the ρ meson $\Gamma(\rho \to \pi\gamma)$, using a dispersive framework.[9,39] The preliminary COMPASS chiral anomaly amplitude is[9] $A_{3\pi} = 10.3 \pm 0.1_{stat} \pm 0.6_{syst}$ $GeV^{-3}$. Ongoing analysis aims to reduce the systematic error, and to include electromagnetic corrections.[9] A recent dispersive analysis of $e^+e^- \to 3\pi$ data yields[59] $A_{3\pi} = 10.0 \pm 0.5$ $GeV^{-3}$. Table 2 summarizes available experimental and theoretical results for $A_{3\pi}$.

Considering uncertainties, the COMPASS preliminary value $A_{3\pi} = 10.3 \pm 0.6$ $GeV^{-3}$ aligns well with both the leading-order (LO) two-flavor ChPT prediction[8,39,58] $A_{3\pi} \approx 9.7 \pm 0.1$ $GeV^{-3}$ and the HO three-flavor ChPT prediction[8,51] $A_{3\pi} \approx 10.7 \pm 1.0$ $GeV^{-3}$, as well as with the dispersive analysis value[59] $A_{3\pi} \approx 10.0 \pm 0.5$ $GeV^{-3}$ of $e^+e^- \to 3\pi$ data. The same can be said for the Serpukhov "value" $A_{3\pi} = 9.6 \pm 1.2$ $GeV^{-3}$. PrimEx studies of the η lifetime[19,20] and the COMPASS AMBER chiral anomaly reactions $\pi\gamma \to \pi\eta$[15] and $\gamma K \to \pi^0 K$[16] are being planned. Combined with further theoretical developments, precision chiral anomaly data are needed for comprehensive tests of three-flavor ChPT.

Table 2. Theoretical and experimental results for $\pi\gamma \to \pi\pi$, with statistical and systematic errors added in quadrature

| Theory(T), Exp(E) | Reference | $A_{3\pi}$ $GeV^{-3}$ |
|---|---|---|
| E, Serpukhov | 52,53,54,55,56 | 9.6 ± 1.2 |
| E, COMPASS | 9 | 10.3 ± 0.6 |
| E, CERN, $\pi e \to \pi\, e\, \pi^0$ | 57, 58 | 9.6 ± 1.1 |
| T, LO | 8, 39, 58 | 9.7 ± 0.1 |
| T, HO | 8, 51 | 10.7 ± 1.0 |
| E+T, $e^+e^- \to 3\pi$ | 59 | 10.0 ± 0.5 |

## 10. Conclusions

This study reviews Primakoff scattering experiments on pion polarizability and the $\pi\gamma \to \pi\pi$ chiral anomaly at CERN COMPASS, as well as measurements of the $\pi^0$ lifetime at Jefferson Lab. Following a brief review of ChPT, we highlight the agreement between these experimental results and predictions from two-flavor (*u, d*) ChPT. This consistency reinforces the identification of the pion as a Goldstone boson arising from spontaneous chiral symmetry breaking. Proposed high-precision Primakoff scattering studies at CERN and JLab are crucial for testing and validating three-flavor (*u, d, s*) ChPT. These include pion and kaon polarizabilities, $\pi\gamma \to \pi\eta$ and $K\gamma \to K\pi^0$ chiral anomaly reactions, and JLab measurements of the η lifetime and pion polarizability.

Incorporating the strange quark, three-flavor ChPT provides a more complete description of the dynamics of light mesons—pions, kaons, and etas. Comparing experimental data for kaons, $\pi^0$ and η with three-flavor ChPT predictions will clarify how well the theory captures the interplay of the additional flavor and the effects of the strange quar. These tests will shed light on the role of pions, kaons, and etas as Goldstone bosons associated with spontaneous chiral symmetry breaking. Such comparisons can identify limitations of three-flavor ChPT, guiding improvements or hinting at new physics.[31,60,61] Ultimately, these comprehensive investigations aim to deepen our understanding of meson dynamics, flavor symmetries, and the implications of chiral symmetry breaking—advancing our knowledge of the strong force and the fundamental nature of particles.

## Acknowledgments

Many thanks to Leonid Frankfurt and Bastian Kubis for helpful suggestions.

**ORCID:** Murray Moinester **-** https://orcid.org/0000-0001-8764-5618; **CRIS:** https://cris.tau.ac.il/en/persons/murray-moinester; **Website**: https://murraymoinester.com